\documentclass{tPHM2e}

\usepackage{graphicx}
\newcommand{\BEQ}{\begin{equation}}
\newcommand{\EEQ}{\end{equation}}
\newcommand{\BEA}{\begin{eqnarray}}
\newcommand{\EEA}{\end{eqnarray}}

\newcommand{\nn}{\nonumber }

\begin{document}
\title{Spin-Glass  Model for Inverse Freezing}

\date{\today}

\author{Luca Leuzzi\thanks{$^\ast$ Email: luca.leuzzi@roma1.infn.it}
$^\ast$
\\
\vspace{6pt} SMC-INFM/Institute of Complex Systems (ISC) CNR, 
\\
and  Department of Physics, 
University of Rome, ``La Sapienza'', Piazzale A. Moro 2, 00185, Rome, Italy}

\maketitle

\begin{abstract}
We analyze the Blume-Emery-Griffiths model with disordered magnetic
interaction displaying the inverse freezing phenomenon. The
behaviour of this spin-1 model in crystal field is studied throughout
the phase diagram and the transition and spinodal lines for the model
are computed using the Full Replica Symmetry Breaking Ansatz that
always yelds a thermodynamically stable phase. We compare the results
both with the quenched disordered  model  with Ising spins on
lattice gas - where no reentrance takes place - and with the model
with generalized spin variables recently introduced by Schupper and
Shnerb [Phys. Rev. Lett. {\bf 93}, 037202 (2004)]. The simplest
version of all these models, known as Ghatak-Sherrington model, turns
out to hold all the general features characterizing an inverse
transition to an amorphous phase, including the right thermodynamic
behavior.
\end{abstract}

\section{Introduction}
In the last years, in a relatively broad 
number of materials an apparently  weird kind of phase
transition has been detected and investigated: the ``inverse melting''.
Such a phenomenon was, actually, already hypothesized
by Tammann a century ago\cite{Tammann} but it has now been 
found experimentally in
polymeric and colloidal compounds,  
high-$T_c$ superconductors,  proteins,  ultra-thin films,  liquid crystals and
metallic 
alloys\cite{RHKMM99,SDTJPC01,CACPS97,PVPNat03,TSYJPS01,PlazanetJCP04,Angelini06}.
This kind of transition, includes, e.g., the solidification
of a liquid or the transformation of an amorphous solid into a crystal upon
heating. The reason for this counter intuitive process is that a phase usually
at higher entropic content happens to exist in very peculiar patterns such
that its entropy is decreased below the entropy of the phase usually
considered the most ordered one.  An example taking place in the widely
studied Crystalline Poly(4-methylpentene-1) (P4MP1)\cite{RHKMM99} 
is the one of a crystal state of
{\em higher} entropy that can be transformed into a fluid phase of {\em lower}
entropy on cooling. Inverse transitions, in their most
generic meaning (i.e. both thermodynamic or occurring by means of kinetic
arrest), have been observed between fluid and crystal phases\cite{SDTJPC01},
between glass and crystal \cite{RHKMM99} and between fluid
and glass (``inverse freezing'')\cite{CACPS97}. For a comprehensive summary of 
materials undergoing some kind of phase transition one can turn to Ref. 
\cite{SS05}. 
The aim of this work is to study a simple mean-field model for the
inverse transition in a spin-glass, in order to heuristically
represent the inverse fluid-amorphous transition.

\section{The Random Blume-Emery-Griffiths-Capel Model}
  We have been analyzing the
Blume-Emery-Griffiths-Capel (BEGC) model with quenched disorder using
the Full Replica Symmetry Breaking (FRSB) scheme of computation that
yields the exact stable thermodynamics\cite{CLPRL02,CLPRB04,CLPRL05}.
It includes the Blume-Capel \cite{CapelPhys66} and the
Blume-Emery-Griffiths\cite{BEGPRA71} models, when the couplings are
ferromagnetic, and the Ghatak-Sherrington (GS) model,\cite{GSJPC77}
when the couplings are random variables and no biquadratic interaction
occurs.

The model  Hamiltonian we consider is 
\BEQ 
{\cal H}=-\sum_{ij}J_{ij}S_iS_j+D\sum_{i=1}^NS_i^2
-\frac{K}{N}\sum_{i<j}S_i^2S_j^2
\label{Hamiltonian}
\EEQ 
where $S=1,0,-1$, $D$ is the crystal field, $J_{ij}$ are quenched
random variables (Gaussian) of mean zero and variance $1/N$.  The
parameter $K$ represents the strength of the biquadratic interaction.
A generalized
spin variable has been recently proposed
\cite{SSPRL04,SS05}, for which
the degeneracy of the magnetically interacting sites ($S=\pm 1$)
can be  larger (or smaller) than the
one of the  holes ($S=0$).  
To be as inclusive as possible we have been looking at
the phase diagram of the random BEGC model also in terms of these generalized
 variables
considering the thermodynamically stable
spin-glass obtained by means of the FRSB Ansatz.

Let us call
call ${k}$ the degeneracy of the filled in sites of one type ($S=1$ or
$S=-1$) and ${l}$ the degeneracy of the empty sites ($S=0$). The
relevant parameter is $r={k}/{l}$.
When $r=1$ the {spin-1} model is
obtained. If, furthermore, $K=0$, the model is the GS one.  When,
otherwise, $r=1/2$ and $D\to\mu= -D$ the lattice gas formulation of
Ref.  \cite{ANSJPF96} is recovered, for which no reentrance was
observed\cite{CLPRL02,CLPRB04}. 

\begin{figure}[t!]
{\includegraphics[width=.48\textwidth]{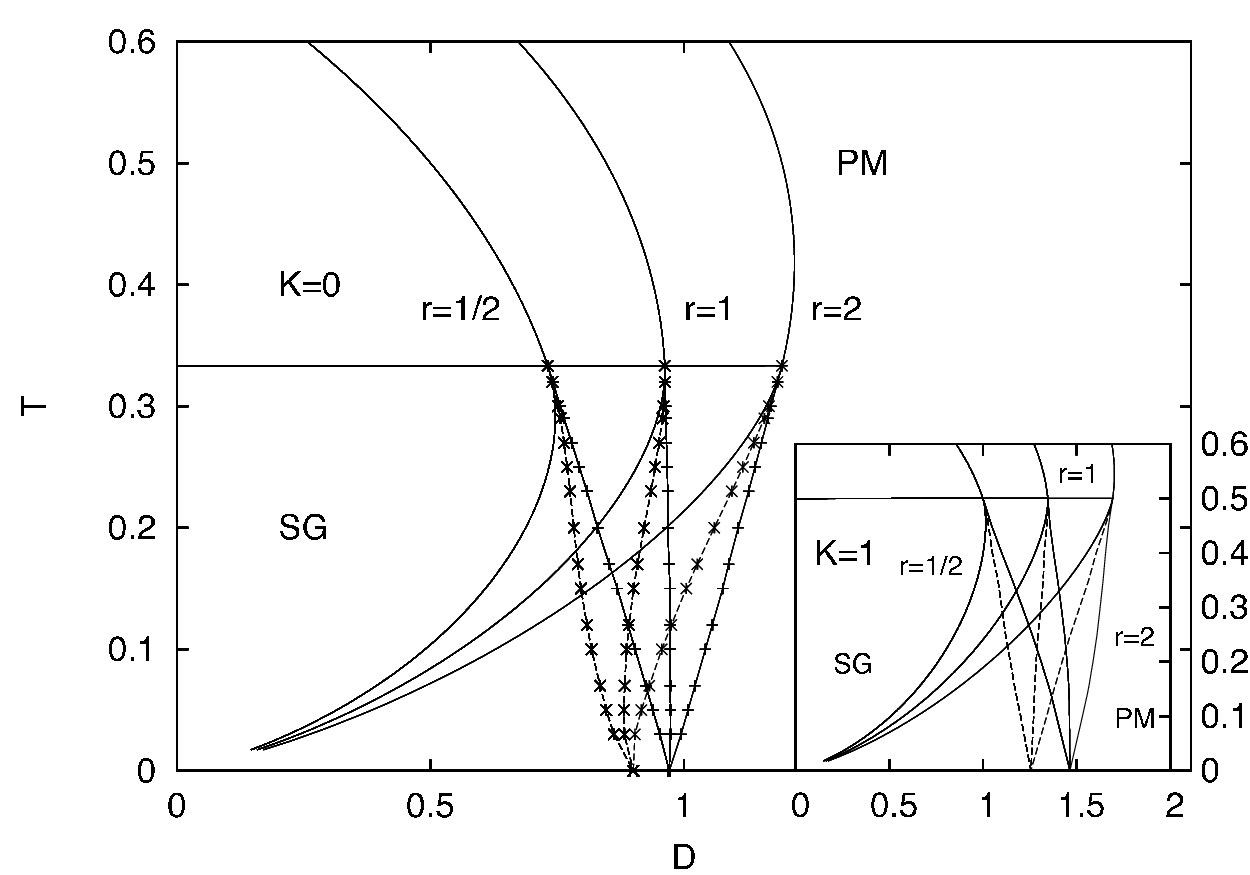}
\includegraphics[width=.48\textwidth]{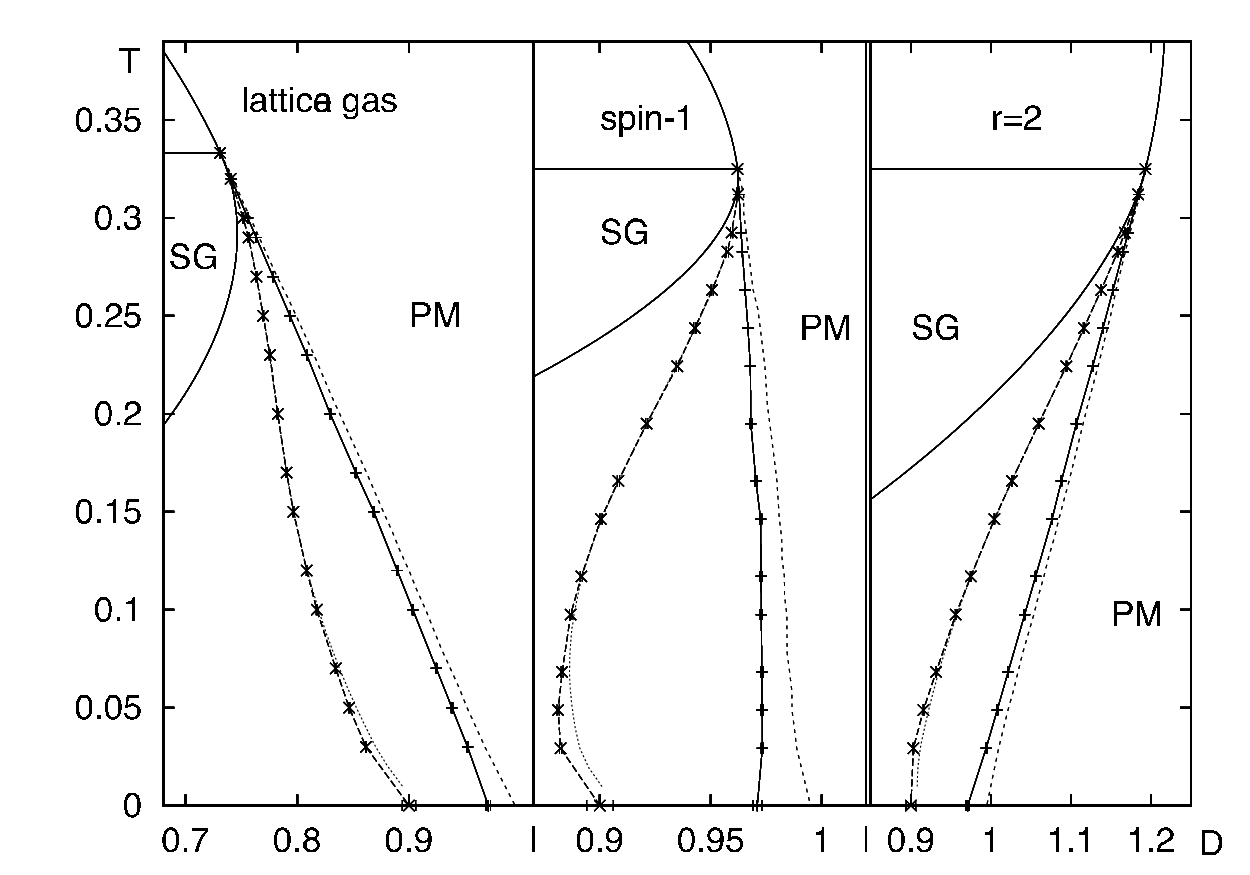}}
\caption{The $D$-$T$ phase diagram in absence of biquadratic
  interaction.  Three models with different behaviors are plotted:
  $r=1/2$, $1$, $2$.  For each $r$-model three curves are represented,
  each departing from the same tricritical point: the full curve on
  the left is the spinodal of the PM phase, the dashed one in the
  middle is the first order transition line and the right one is the
  SG spinodal line.  The group of three curves on the left are for the
  Ising spins on lattice gas ($r=1/2$, $T_c=1/3$, $D_c=0.73105$). The
  group of curves in the middle represent the lines of the GS model
  ($r=1$, $D_c=0.96210$). The curves on the right correspond to the
  $r=2$ model ($D_c=1.19315$).  In the inset the same diagram is plotted
  when $K=1$ (attractive biquadratic interaction). No qualitative difference 
occurs.}
\label{fig:phdiK0}
\caption{$D$-$T$ diagrams at $K=0$:
  Ising spin glass on lattice gas (left), GS model (center) and
  random BEGC model with $r=2$ (right).  Both the RS and the FRSB
  solution (the latter with error bars) are plotted. For $r=1/2$ no
  reentrance occurs. A reentrance occurs,
  instead, below the tricritical point in the GS model along the first
  order transition line (and a second reentrance seems is there for
  $T<0.03$). In the latter model 
  the reentrance occurs above the tricritical point.}
\label{fig:rdet}
\end{figure}

We discuss the physically stable solution for both the GS
model\cite{GSJPC77,MSJPC85} ($K=0$) and the model with attractive biquadratic
interaction\cite{CLPRB04} ($K/J=1$), whose phase diagrams are plotted in Fig.
\ref{fig:phdiK0}, and we compare it with the Replica Symmetric (RS) results (see Fig.
\ref{fig:rdet}).  We study the behaviour of the phase diagram for (a) the
lattice gas case ($r=1/2$), for which no inverse transition occurs anywhere in
the parameter space; (b) the spin-1 case ($r=1$), where the first order
transition line displays a reentrance soon below the tricritical temperature;
(c) the generalized cases as $r>1$, in particular we plot the results of the
model with variables taking values $S=\{1,1,0,-1,-1\}$ for which the reentrance
takes place above the tricritical point, along the second order phase
transition line.

\section{The Full Replica Symmetry Breaking scheme of computation for the Random
BEGC model}
Using the replica trick\cite{MPV} the free energy of $n$ replicas of the system
with Hamiltonian Eq. (\ref{Hamiltonian}) turns out to be
\BEA
\label{f_rep}
&& n \beta f = \frac{\beta}{2}\left(K+\frac{\beta}{2}\right)\sum_{a=1}^n\rho_a^2
+\frac{\beta^4}{4}\sum_{a\neq b}q_{ab}^2
-\log \sum_{\{S\}}\exp(-\beta {\cal H}')  
\\
&& -\beta {\cal H}'=\beta \left(K+\frac{\beta}{2}\right)
\sum_{a=1}^n\rho_aS_a^2
-\beta D\sum_{a=1}^nS_a^2+\frac{\beta^2}{2}\sum_{a\neq b}q_{ab}^2 S_aS_b
\EEA
where the order parameters introduced in the computation satisfy the 
self-consistency equations
\BEQ
\label{self_con}
\rho_a=\left<S_a^2\right>\hspace*{2cm}
q_{ab}=\left<S_aS_b\right>
\EEQ

Choosing the Parisi Ansatz for the matrix $q_{ab}$ and
performing infinite steps of 
RSB\cite{ParisiJPA80} the free energy of the spin-glass phase appears to be
\BEQ
\label{f_frsb}
\beta f = \frac{\beta }{2} \left(K+\frac{\beta}{2}\right)
\rho^2-\frac{\beta^2}{4}\int_0^1 dx~q^2(x)
-\beta\phi(0,0)
\EEQ
and the self-consistency equations become
\BEQ
\label{self_con_frsb}
q(x)=\int_{-\infty}^\infty dy~P(x,y) \phi'(x,y)^2
\hspace*{2cm}
\rho = \int_{-\infty}^\infty dy~P(1,y)\frac{2 r e^\Theta\cosh \beta y}
{1+2 r e^\Theta\cosh \beta y}
\EEQ
where $\Theta\equiv \beta^2/2[\rho-q(1)]+\beta K\rho-\beta D$.
The functions $\phi(x,y)$ and $P(x,y)$  are solutions of the 
non-linear
partial differential equations
\BEA
\label{Parisi_eq}
&&\dot\phi(x,y)=-\frac{\dot q}{2}\left\{\phi''(x,y)+\beta x \left[
\phi'(x,y)\right]^2\right\}
\\
&&\dot P(x,y)=-\frac{\dot q}{2}\left\{
P''(x,y)-2\beta x\left[P(x,y)\phi'(x,y)\right]'\right\}
\EEA
with boundary conditions
\BEQ
\phi(1,y)=\frac{1}{\beta}\log\left(2+4 r e^\Theta\cosh \beta y\right)
\hspace*{2 cm} P(0,y)=\delta(y)
\EEQ
The dot and the apex respectively represent the derivative with respect to $x$  
and $y$. For a detailed study of the above equations the reader can look
 at Ref. \cite{CLPRB04}.

Having a model with variables displaying a relative degeneracy $r$
($D=D_r$), in order to describe the partition function of another model
whose variables have degeneracy $r'$ 
it is enough to vary the crystal field as
$ 
D_{r' }+ T\log r' = D_{r} +
T\log r
$. 
This does not hold, however, for the state
functions obtained deriving the thermodynamic potential 
with respect to the temperature
(e.g. entropy and internal energy) 
that will, instead, receive contributions from additional terms.
For instance, the entropy function ($K=0$) is
\BEA
\label{entropy}
s(\beta,D)&=&-\frac{\beta^2}{4}\left[\rho-q(1)\right]^2
-\frac{\beta^2}{2}\rho\left[\rho-q(1)\right]
-\beta \rho D +\rho\log(2r)
+\beta^2\int_0^1dx~q^2(x)\\
\nn
&&
+\int_{-\infty}^\infty dy~P(1,y)\log
\left\{2+4\exp\left(\Theta +\log r\right)\cosh\beta y\right\}
\EEA
Identifying $D_{1/2}\equiv -\mu$ in Eq (\ref{entropy}) one  recovers the case of
spins on a lattice gas of chemical potential
$\mu$\cite{CLPRL02}.

\section{Thermodynamics}
By solving  
 Eqs. (\ref{f_frsb})-(\ref{entropy})
we  build the $D$-$T$ phase diagrams that we plot for $K=0$ and $K=1$ in Fig.
\ref{fig:phdiK0} and inset, at different values of $r$. 
The reentrance in the $D$-$T$ plane is present already in the
spin-1 GS model.  As a consequence this implies that there is no need
for the intuition of Ref.  \cite{SSPRL04} in order to
have a model for inverse freezing from low temperature liquid to high
temperature amorphous solid.  This is at difference with the
liquid-crystal inverse transition (``inverse melting'') for the
description of which the original Blume-Capel model is not adequate
and $r>1$ is needed\cite{SSPRL04,SS05}.

The transition lines are
not very much dependent on the Ansatz used to compute the quenched
average of the free energy. For not extremely low $T$, the
first order transition lines yielded by the RS and the FRSB Ansatz
even coincide down to the precision of our numerical evaluation of the FRSB
antiparabolic Parisi equation (\ref{Parisi_eq})
 whereas they slightly differ at very
low temperature (see inset of Fig.  \ref{fig:rdet}). 
For what concerns the
spinodal lines, the RS ones are shifted  inside the
pure paramagnetic (PM) phase.

Introducing a biquadratic interaction term and varying it from
attractive to repulsive the situation does not change much (see
e.g. inset of Fig. \ref{fig:phdiK0}). 
For any value of  $K$ no
reentrance of the phase transition line occurs in the $D$-$T$ phase
diagram of the lattice gas model whereas it is {\em always} there for
the spin-1 model. The only consequence of reducing $K$
 is that the area of the phase coexistence region is
reduced (the tricritical temperature tends to zero as $K\to-\infty$).

The slope of a first order line is given by the Clausius-Clapeyron
equation.  For the BEGC model it can be written in terms of the crystal
field $D$ (playing the role of a chemical potential), instead of the
pressure that is not defined in our model:\footnote{With respect to the classic
Clausius-Clapeyron equation $D$ takes the place of the pressure and
$\rho$ plays the role of the specific volume.}
\BEQ
\frac{d D}{d T}=\frac{s_{\rm pm}-s_{\rm sg}}{\rho_{\rm pm}-\rho_{\rm sg}}
=\frac{\Delta s}{\Delta\rho}
\label{CC}
\EEQ 
This formula is valid for any $r$. We stress that in passing from
$r$ to $r'$ also the entropy (Eq. \ref{entropy}) 
changes of a term $\rho \log r/r'$, in
agreement with the crystal field shift given above.

\begin{figure}[t!]
{\includegraphics[width=.49\textwidth]{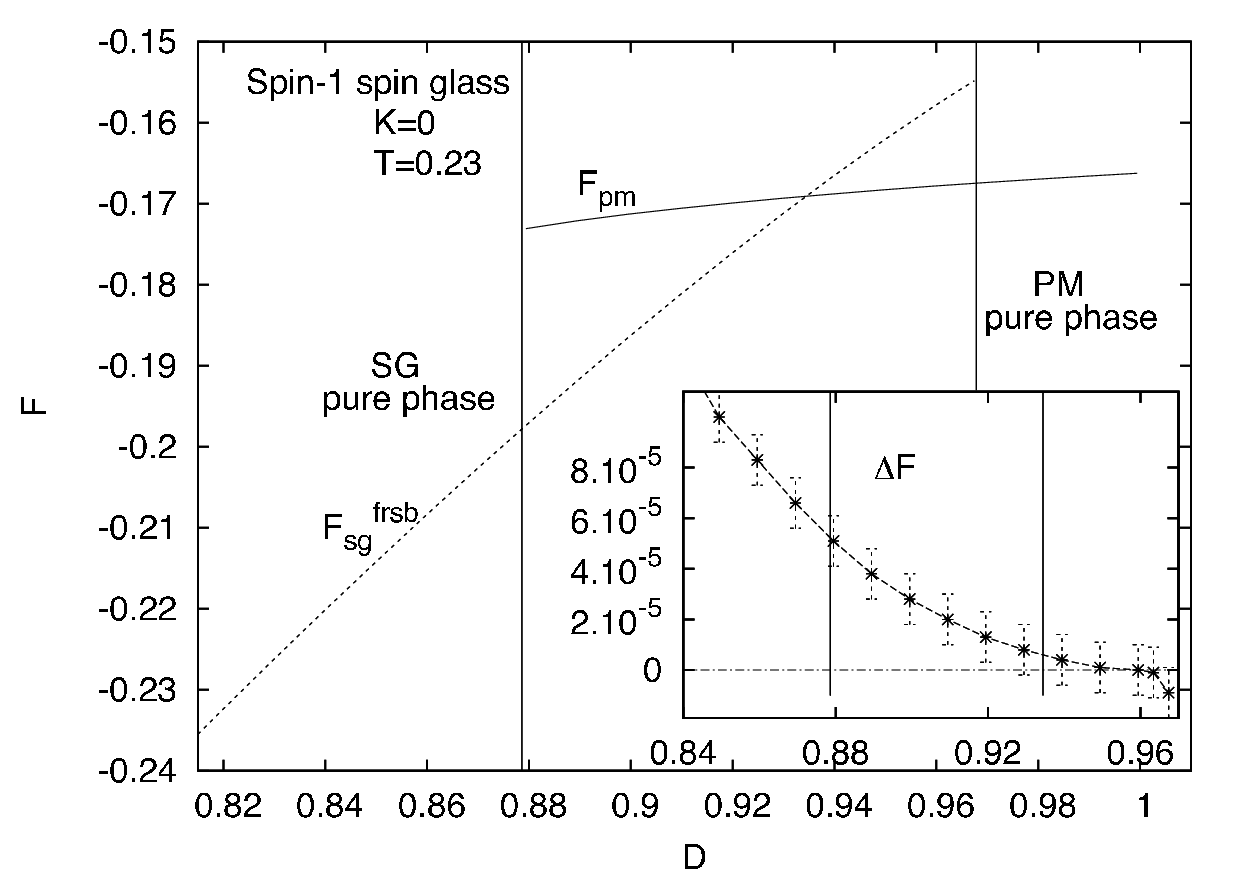}
\includegraphics[width=.49\textwidth]{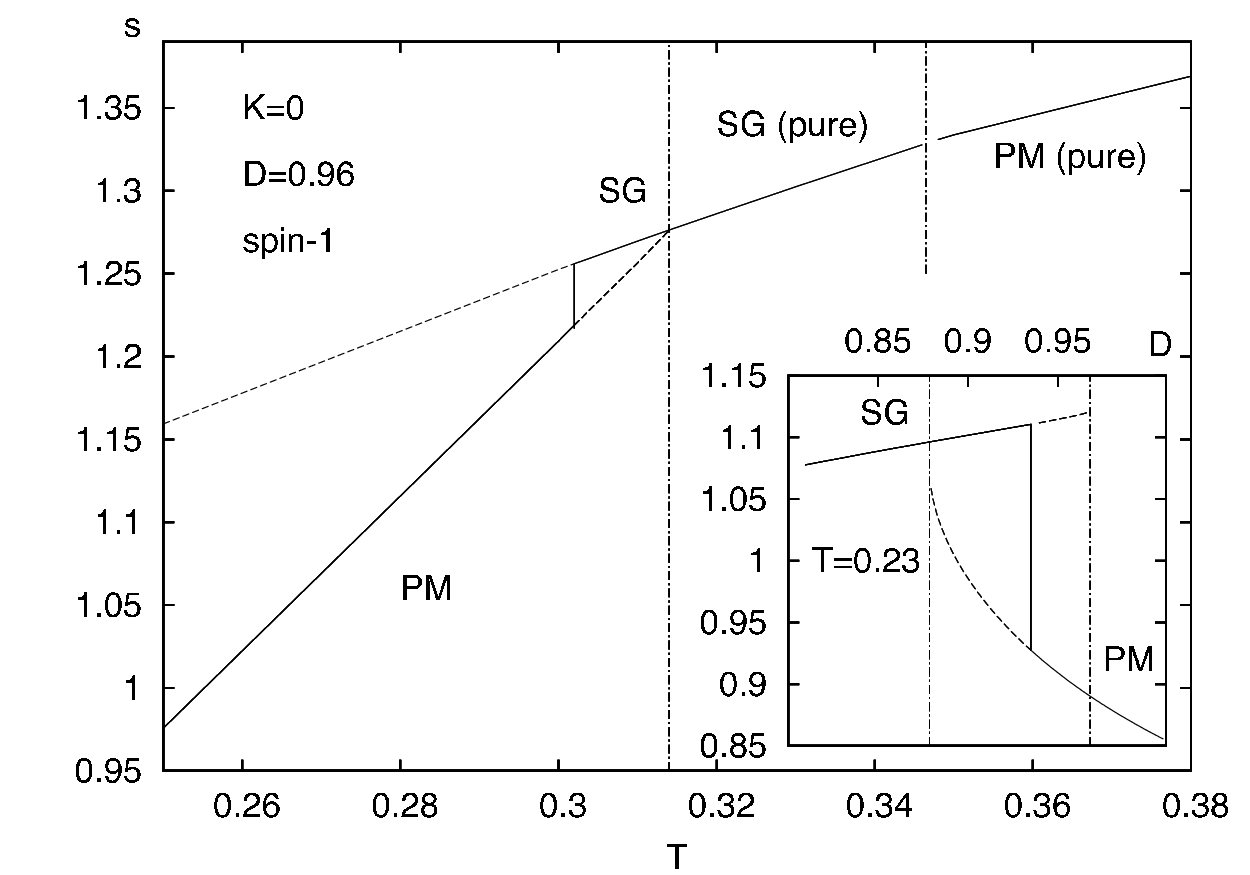}}
\caption{Free energy $F_{\rm pm}$ and $F_{\rm sg}^{\rm frsb}$
  vs. $D$ at $T=0.23$ in the GS model
  ($K=0$). The left and right side vertical lines are
  at the spinodal points of, respectively, the
  PM and the SG phase. 
  The first order transition occurs at $D_{1^{st}}=0.9344$. In
  the inset $\Delta F=F_{\rm sg}^{\rm frsb}-F_{\rm
  sg}^{\rm rs}$ is displayed. At this temperature the two functions
  merge very near to the tricritical point (right side vertical line).}
\label{fig:F_D}

\caption{Entropy vs. $T$ at the crystal field value of $D=0.96$ for
the GS model. $T_{1^{st}}=0.302$. The two vertical dashed lines are
the SG ($T_{\rm SG}=0.314$) and PM ($T_{\rm PM}=0.3465$) spinodal
lines.  In the inset $s(D)$ is plotted at $T=0.23$.}
\label{fig:entro}
\end{figure}

In Fig. \ref{fig:entro} the behaviour of the entropy as a function of the
temperature (Eq. (\ref{entropy}))
is shown across an inverse transition (as a function of
the crystal field $D$ in the inset) for the spin-1 model.  The entropy
of the PM phase {\em below} the first order transition line is {\em
smaller} than the entropy of the SG: heating the system the paramagnet
becomes an amorphous magnet (i.e. ``freezes'') acquiring latent heat
from the heat bath.
Going down along the transition line, as $\Delta s$ changes sign in Eq. 
(\ref{CC}) the
slope becomes positive. The $\Delta s=0$ point is called a Kauzmann
locus\cite{SDTJPC01}.

A reentrance in the transition line
can be due both to the existence of a liquid phase with an entropy
lower than the one of the solid or to a the liquid phase more
dense than the solid one (like in the water-ice transition).  When an
entropic inversion accounts for the phase transition, the
equilibrium transition line changes slope in a point where the entropy
of the fluid phase, $s_{\rm liq}$, becomes equal to the one of the solid, 
$s_{\rm sol}$, according to
the Clausius Clapeyron equation for first order phase transitions.
From this point on a whole iso-entropic line, $\Delta s = s_{\rm liq}
-s_{\rm sol}=0$, can
be continued both inside the solid and the liquid phases.
This is a particularly interesting observation since, in the context
of glass formers, a
transition to an ``ideal'' glass at the temperature at which $\Delta
s=0$ is hypothesized, 
in order to avoid the paradox that an under-cooled liquid might
possess less entropy than the associated crystal at the same values of
the thermodynamic parameters\cite{Kauz48}.  From an experimental point of view 
this Kauzmann temperature would be the temperature of the glass transition
(that is not a true phase transition because strictly kinetic in
origin) in an idealized adiabatic cooling procedure. Since the
astronomically long relaxation time needed to actually perform such an
experiment makes such a procedure unfeasible, the evidence in favour of
the existence of a thermodynamic glass transition mainly comes from
analytical and numerical investigations (see
e.g. Refs. \cite{MPJCP,CPV1}).  The fact that a
$\Delta s=0$ line turns out naturally in the description of the
behavior of materials with inverse transition avoids, at least for
these substances, the Kauzmann paradox and breaks the connection
between $\Delta s=0$ extrapolation and the existence of an ideal
amorphous phase\cite{SDTJPC01,RHKMM99}. 

\section{Conclusions}
We have shown that the Ghatak-Sherrington model
undergoes the inverse freezing phenomenon acquiring
latent heat from the heat bath as the paramagnet becomes a spin-glass.
Many other
models can be built starting from this one, introducing an attractive
or repulsive biquadratic interaction (the last term in Hamiltonian
(\ref{Hamiltonian})) and/or tuning the relative degeneracy of the value $S=0$
and $S^2=1$ of the spin variable (as in \cite{SSPRL04}) 
but the GS one already contains 
all the features needed to qualitatively
represent the experimental results.

We have considered here 
a case where the frozen phase is a spin-glass and the transition 
is only static: in order to have a structural glass and to deal with a 
dynamic (mode coupling) transition, 
it would be, however, enough to 
apply the simpler
one step RSB Ansatz in, e.g., a model with
a $p$-body magnetic interaction (with $p>2$ instead of a couple interaction),
 such as the one introduced in Ref. \cite{MottishawEPL86} 
or an analytically solvable  model with
spherical variables instead of discrete ones and again a multi-body interaction
(a valuable starting point would be the work of 
Ref. \cite{CCNPRL04}).
Else, a spin-1 model with an orthogonal matrix $J_{ij}$, 
instead of the Gaussian one, can be used, as latterly done by Sellitto
\cite{Sellitto06}.
Besides the inverse freezing phenomenon this last model also displays a 
fluid-fluid transition (between two distinct paramagnetic phases), a feature 
recently found in the solution of 4-methil-piridyne (4MP) and 
$\alpha$-cyclodextrin in water, investigated
by Angelini and coworkers\cite{Angelini06}.

With the help of the class of models presented here
 the connection between entropy driven phase reentrance and shear {\em thickening}
 can also  be tackled\cite{SKPRL05}
 and, furthermore, 
a generalization of the spin-1 variable to a composition of
 ``fast'' and ``slow'' variables \footnote{E.g., setting $S=\sigma n$, 
with $n=0,1$ fast 
and $\sigma = \pm 1$ slow} coupled to two different thermal baths 
 allows for studying anomalous latent heat in out of equilibrium transitions
\cite{APPRL06}.

\end{document}